\documentclass[showpacs,twocolumn,preprintnumbers,amsmath,amssymb]{revtex4}
\usepackage{mathrsfs}
\usepackage{graphicx}
\usepackage{dcolumn}
\usepackage{bm}
\newcommand{\p} {\partial}
\begin{document}
\title{The transport mechanism of  the integer quantum Hall effect}
\author{W. LiMing}\email{wliming@scnu.edu.cn}\author{Tan hui}
\begin{abstract}The integer quantum Hall effect is analysed using a transport mechanism  with a semi-classic wave packages of electrons in this paper. A strong magnetic field perpendicular to a slab separates the electron current into two branches with opposite wave vectors $({\it k})$ and locating at the two edges of the slab, respectively, along the current. In this case  back scattering of electrons ($k\rightarrow -k$)  is prohibited by the separation of electron currents. Thus the slab exhibits zero longitudinal resistance and plateaus of Hall resistance. When the Fermi level is scanning over a Landau level when the magnetic field increases, however, the electron waves locate around the central axis of the slab and overlap each other thus back scattering of electrons takes place frequently. Then longitudinal resistance appears and the Hall resistance goes up from one plateau to a new plateau.
\end{abstract}
\date{\today}
\maketitle \indent

\section{introduction}
Klitzing {\it et al} discovered the integer quantum Hall effect(IQHE) in 1980\cite{Klitzing}.
The classic Hall effect on a slab exhibited  a quantization phenomenon in a mesoscopic two-dimensional system in a strong magnetic field at a low temperature. Plateaus with integer times of ${e^2/h}$ appear on the curve of the Hall electric conductance with respect to the magnetic field and in addition the longitudinal resistance vanishes at the positions of the plateaus. Two years later Tsui {\it et al} observed the fractional quantum Hall effect\cite{Tsui}, where the Hall conductance showed new plateaus with fractional times of ${e^2/h}$. These striking  quantization effects attract great attention of the scientific world. Some new fields, such as the topological insulators, have been set up upon the quantum Hall effect. In 2013 Xue {\it et al} discovered an anomalous quantum Hall effect that the quantum Hall conductance appears in a ferromagnetic material without any external magnetic field\cite{Xue}.

Since the discovery of the IQHE some theoretical explanations have soon been proposed for the plateaus of the Hall conductance. A widely adopted point of view is that the Landau levels of the slab in a strong magnetic field are broadened into a Gaussian shape due to impurities and disordering according to the theory of Anderson localization of disordered systems. The central part of a broadened Landau level contains extended states and other states in the edges are localized states originating from impurities and disorder. When the Fermi level sweeps the localized states at certain strengths of the magnetic field the Hall conductance remains unchanged since the electrons in the localized states has no contribution to the conductance (thus a plateau appears!)\cite{Prange,Ni}. It will be shown, however, that this picture is not true in Sect II. Due to the confinement of a finite slab the Landau levels are no more straight lines.

Another explanation is the topological view that a Hall plateau of the IQHE is concerned to a TKNN number(A Chern number)\cite{TKNN} based on the Kubo formulas of the linear response of electrons. This theory was first proposed  for a sample with an infinite size and was then generalized to a finite sample to include edge states\cite{Yasuhiro Hatsugai}. 
The problem is that the relation between the Hall conductance and the TKNN number is not very transparent and includes some stringent requirements.
In addition the longitudinal resistance of the IQHE was not considered in this mechanism.

Supriyo Datta proposed an idea of electron transport for the IQHE\cite{Datta}. In a strong magnetic field electrons with high momenta in a two dimensional material are almost localized in the  edges of the sample but the energy of electrons goes up from the Landau level energies. Due to collisions between electrons and the lattice electrons become semi-classic packages with a group velocity given by $v={1\over \hbar}{\p E\over \p k}$. The most important is that the electron currents moving in two opposite directions deviate in space and locate at the two edges of the sample  thus electrons cannot be scattered from one state to another state with an opposite momentum. Thus the longitudinal resistance vanishes and the Hall resistance remains unchanged. This idea provides an elegant explanation to the Hall plateaus of IQHE.

In the following sections we first use Datta's idea to calculate the Hall plateaus of the IQHE. And then we analyse the longitudinal resistance according to the Boltzmann equation on the electron collision.

\section{Electron transport}
Consider an infinitely long two-dimensional slab along the $x$-axis in a uniform magnetic field ${\bf B}$ perpendicular to the the slab. Choose a gauge potential ${\bf A} = (-By, 0)$ for the magnetic field. The Hamiltonian of an electron in the slab is given by
\begin{align}
  H={1\over 2m}[(-i\hbar \partial_x -eBy)^2 - \hbar^2 \partial_y^2] + V_c(y)+ V_h(y)
\end{align}
where $V_c(y)$ is the confining potential of the slab on the two sides,  and $V_h$ is the electric potential in the slab due to the Hall effect. The confining potential $V_c$ is simply set to be $0$ inside the slab and $\infty$ at the two boundaries $y=-{W\over 2}, {W\over 2}$, here $W$ is the width of the slab. The electric potential $V_h = e{U_h\over W}y$, here $U_h$ is the Hall voltage across the slab.

Since the Hamiltonian is independent on $x$ it has eigen-functions in the form,
$ {1\over \sqrt L} e^{ikx} \chi(y)$, here $L$ is the length of the slab and $\chi_{nk}(y)$ satisfies the following dimensionless equation
\begin{align}
\big[( k -2\pi\alpha y)^2 - \partial_y^2 + v_h(y)\big]\chi_{nk}(y)= E_{nk}\chi_{nk}(y)
\end{align}
In the above equation the dimensionless $y$ is the times of a given width $a\sim 1\mu$m.  The dimensionless energy $E_{nk}$ and the Hall potential $v_h$ have got rid of an energy unit $t={\hbar^2\over 2m a^2}$, and the magnetic field $B$ is given by the dimensionless constant $\alpha = {eBa^2\over h}$.

The wave functions and the Landau levels  are computed by solving this equation numerically. The lowest four Landau levels $E_n(k)$ and the modulus of the wave functions of the second Landau level are shown in Fig.1(a,b). The Hall potential causes the Landau levels inclining slightly. Due to the confinement of the two boundaries of the slab the levels rise up rapidly at larger wave vectors. The wave functions with harge {\it k} are localized on the boundaries of the slab as seen in Fig.1(b).

\begin{figure}
\includegraphics[width=1.8in,height=1.8in]{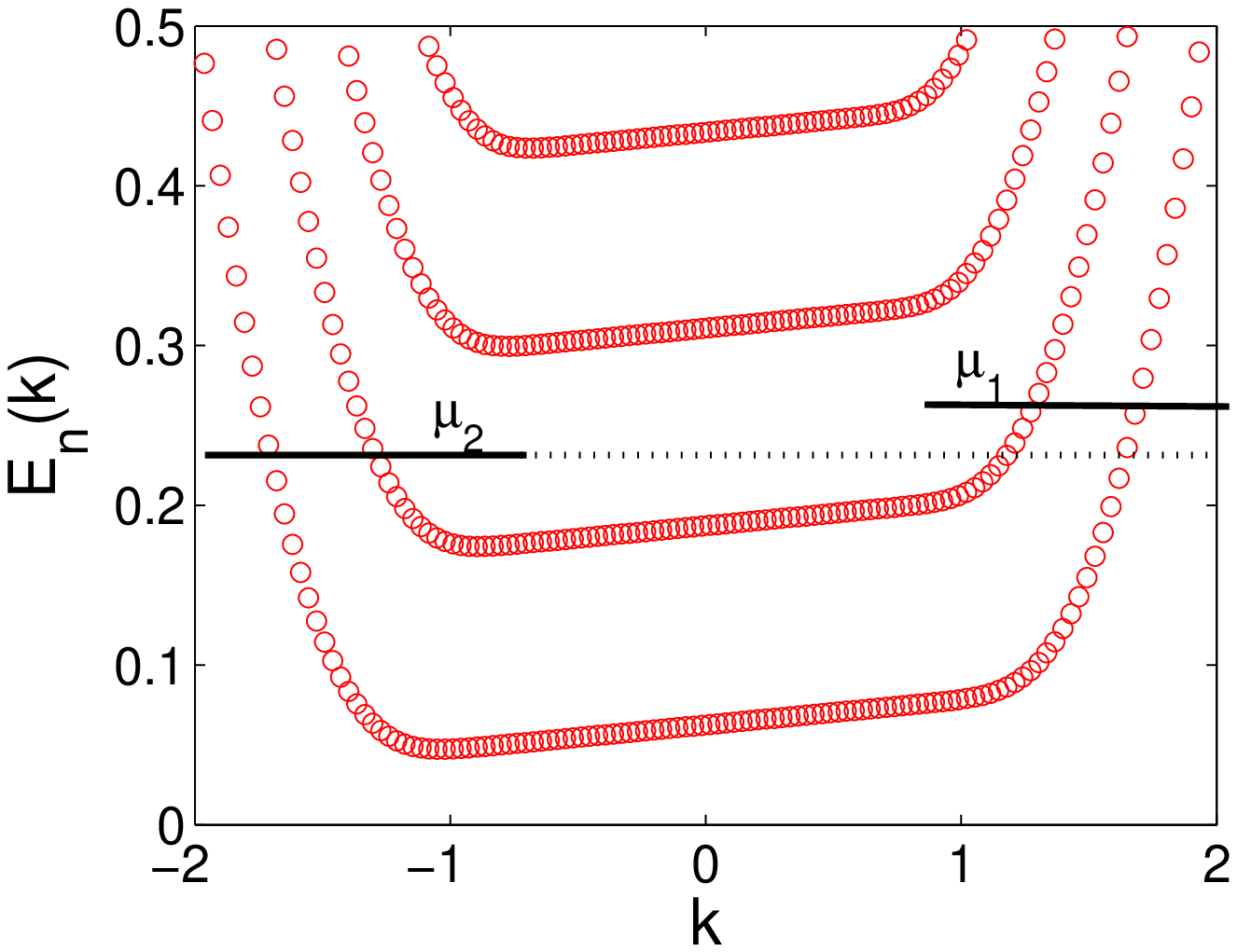}\put(-100,100){(a)}\includegraphics[width=1.8in,
height=1.8in]{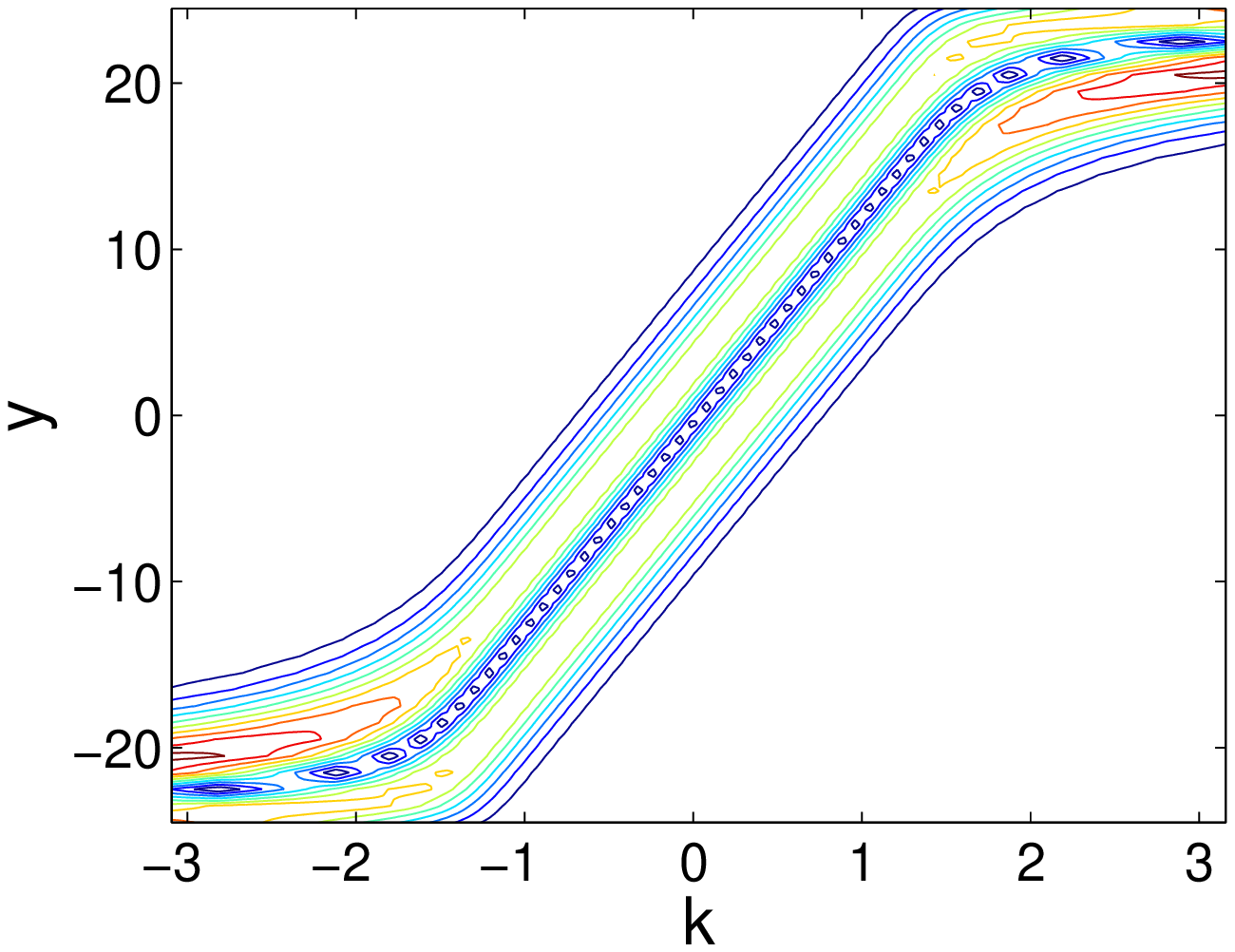}
\put(-100,100){(b)}
\caption{(a) Lowest four landau levels and (b)the modulus of the wave functions of the second Landau level of a slab of width $W=50$ with a Hall potential $v_h=0.05$ in a magnetic field $\alpha =0.01$.  $\mu_1$ and $\mu_2$ are the chemical potentials of electrons with wave vectors $k>0$ and $k<0$,respectively.}
\label{landau}
\end{figure}

 Electrons in crystals in general move as half-classic objects,  wave packages. The group velocity of a wave package is given by the slope of the energies, i.e.
\begin{align}
  v_n(k)={1\over \hbar} {d E_{nk}\over d k}\label{velocity}
\end{align}
This group velocity gives the electric current in the x-axis
\begin{align}
I = -e \sum_n \int {dk\over 2\pi} v_n(k) f(E_{nk})\label{current}
\end{align}
where $f(E)$ is the Fermi-Dirac distribution function. The degeneracy of the spin freedom of electrons has been removed by the strong magnetic field. A different chemical potential for $k>0$ from that for $k<0$ is necessary to give a net current in the slab. Substituting the group velocity into the above integral results in
\begin{align}
I &=-e \sum_n\Big[ \int_{k<0}... +\int_{k>0}... \Big]\nonumber \\
&=-{e\over h} \sum_n\Big[ \int_{\infty}^{E_{n0}}dE_k f(E_k,\mu_2) +\int^\infty_{E_{n0}}dE_k f(E_k,\mu_2) \Big]\nonumber \\
&\approx -M{e\over h} (\mu_1-\mu_2)  = M{e^2\over h} V, M = 0,1,2,...
\label{quantum}
\end{align}
where $M$ is the number of Landau levels cut by the two chemical potentials, $E_{n0}$ is the energy at $k=0$ of the $n$th Landau level, $V$ is the voltage applied to the slab and $-eV = \mu_1-\mu_2$. It is seen that the conductance is quantized with a quanta of ${e^2\over h}$, just the same as the quantization of the IQHE. The above result shows that the current depends not on the detail but the number of the energy bands.

In fact van Wees {\it et al.} observed a quantization with a quanta of ${2e^2\over h}$ on a waveguide with respect to a gate voltage across the waveguide in 1988\cite{Wees} at weaker magnetic fields. The factor 2 of this quanta is due to the spin degeneracy at weak magnetic fields. This quantization is quite similar to the IQHE, however, instead of the gate voltage the IQHE has a Hall voltage.

Datta argued that the IQHE can be indeed explained by such quantization. When the longitudinal resistance vanishes the chemical potentials on the two boundaries remain from the leads\cite{Datta} as seen in Fig.2(a). In this case 
a Hall voltage appears on the two boundaries of the slab due to the difference of the chemical potentials, {\it i.e.} $eV_h = \mu_1-\mu_2$. This explanation is exactly the same as eq.(\ref{quantum}). Further, this current distribution has also explained the zero longitudinal resistance at the plateaus of the IQHE since the chemical potential is the same everywhere on each boundary of the slab.

Numerical results of wave functions do support the above picture of electron transport. As seen in Fig.1(b) wave functions with larger $k$ vectors approach to one of the two boundaries of the slab. These electrons have great group velocities according to (\ref{velocity}) thus give great contributions to the current. In particular the electrons with opposite $k$ vectors separate in space.  The wave functions with small $k$ vectors, however, mix in space around the central axis of the slab. This significant difference is the key to understand the IQHE.

\begin{figure}
\includegraphics[width=1.6in,height=1in]{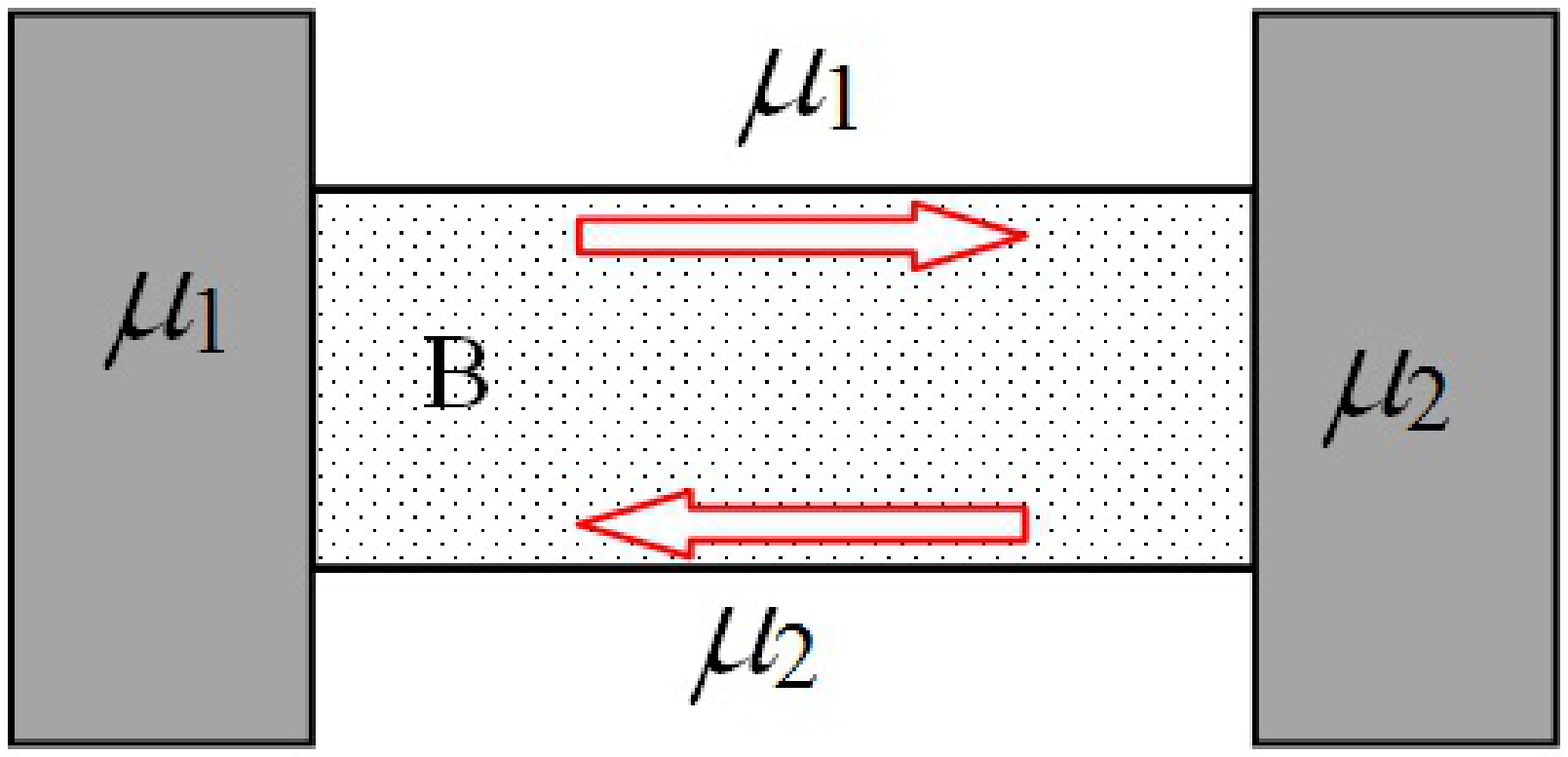}\put(-85,100){(a)}\includegraphics
[width=1.8in,height=1.8in]{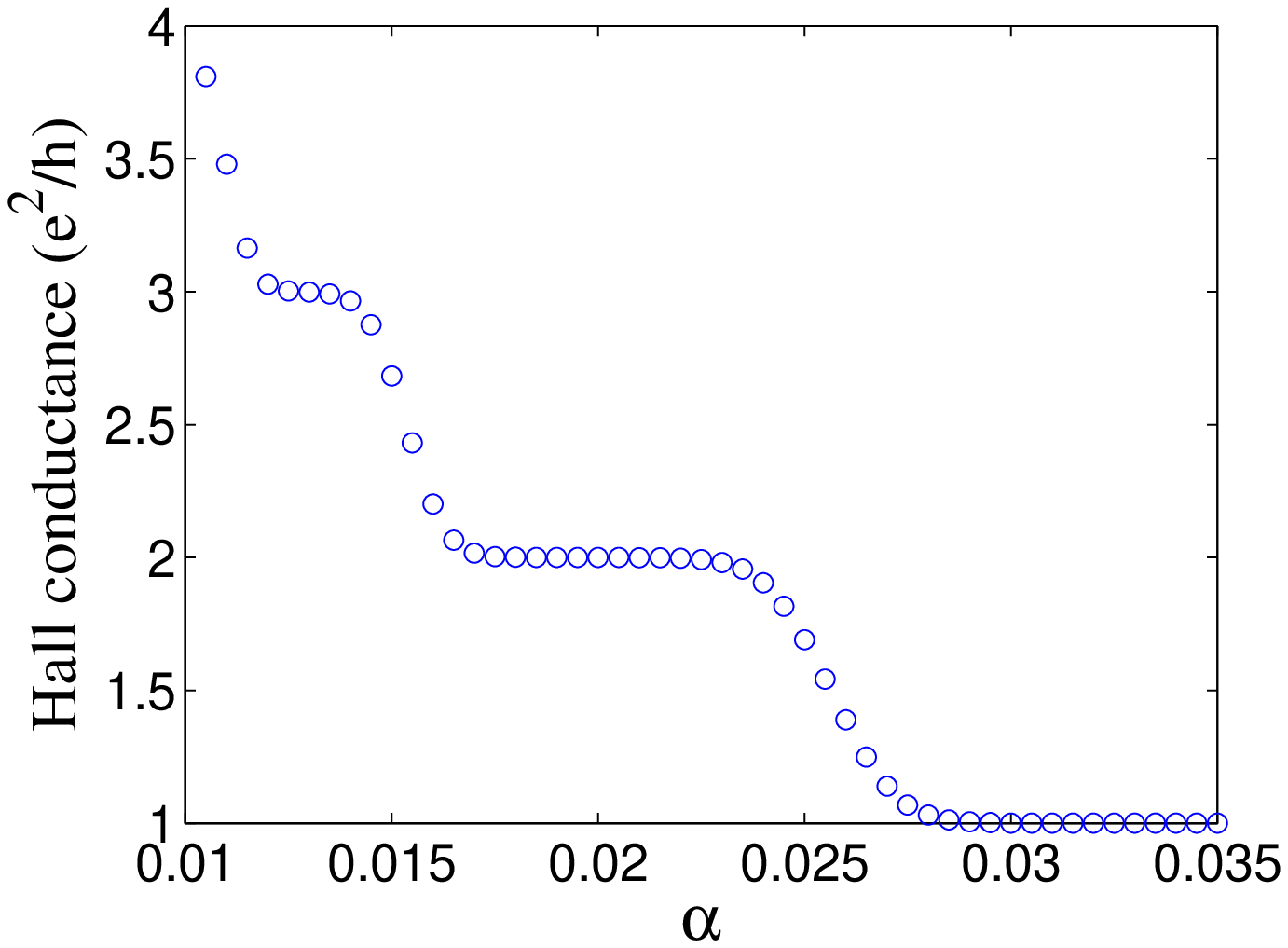}
\put(-90,100){(b)}
\caption{(a)A slab in a magnetic field. Arrows show the electron transport directions. Dots on the slab show the outward direction of the magnetic field.(b)The quantization of the Hall conductance. Parameters see the caption in Fig.1 and $kT=0.01, \mu_1=0.51,\mu_2=0.5$ .}\label{ed}
\end{figure}

The approximation in eq.(\ref{quantum}) is not valid when the Fermi level is close to the bottom of a Landau level. An accurate integration for eq.(\ref{current}) results in
\begin{align}
I&= {e kT\over h} \sum_n \ln{e^{(E_{n0}-\mu_2)/kT}+1\over e^{-eV_h/kT}+e^{(E_{n0}-\mu_2)/kT}}=G_hV
\end{align}
When $E_{n0}\ll\mu_2$ and $kT \ll \mu_2$ the above current approaches to eq.(\ref{quantum}). This accurate result reproduces the plateaus and the transitions between them of the Hall conductance of the IQHE as seen in Fig.2(b). This result assumes a zero longitudinal resistance along the current at both edges of the slab.  This is  true in the plateau regions where the longitudinal resistance indeed vanishes. In the transition regions although the chemical potential drops along the direction of the transport of electrons the Hall voltage remains between the two edges of the slab because of the equilibrium between the electric force and the Lorentz force across the slab. The current may slightly changes due to the drop of the chemical potential in this case. 

\section{Longitudinal resistance}

Next we analyse the longitudinal resistance of the IQHE. Why the longitudinal resistance at the plateaus of the Hall resistance vanishes and goes up between the Hall plateaus?

\begin{figure}
\includegraphics[width=1.8in,height=1.8in]{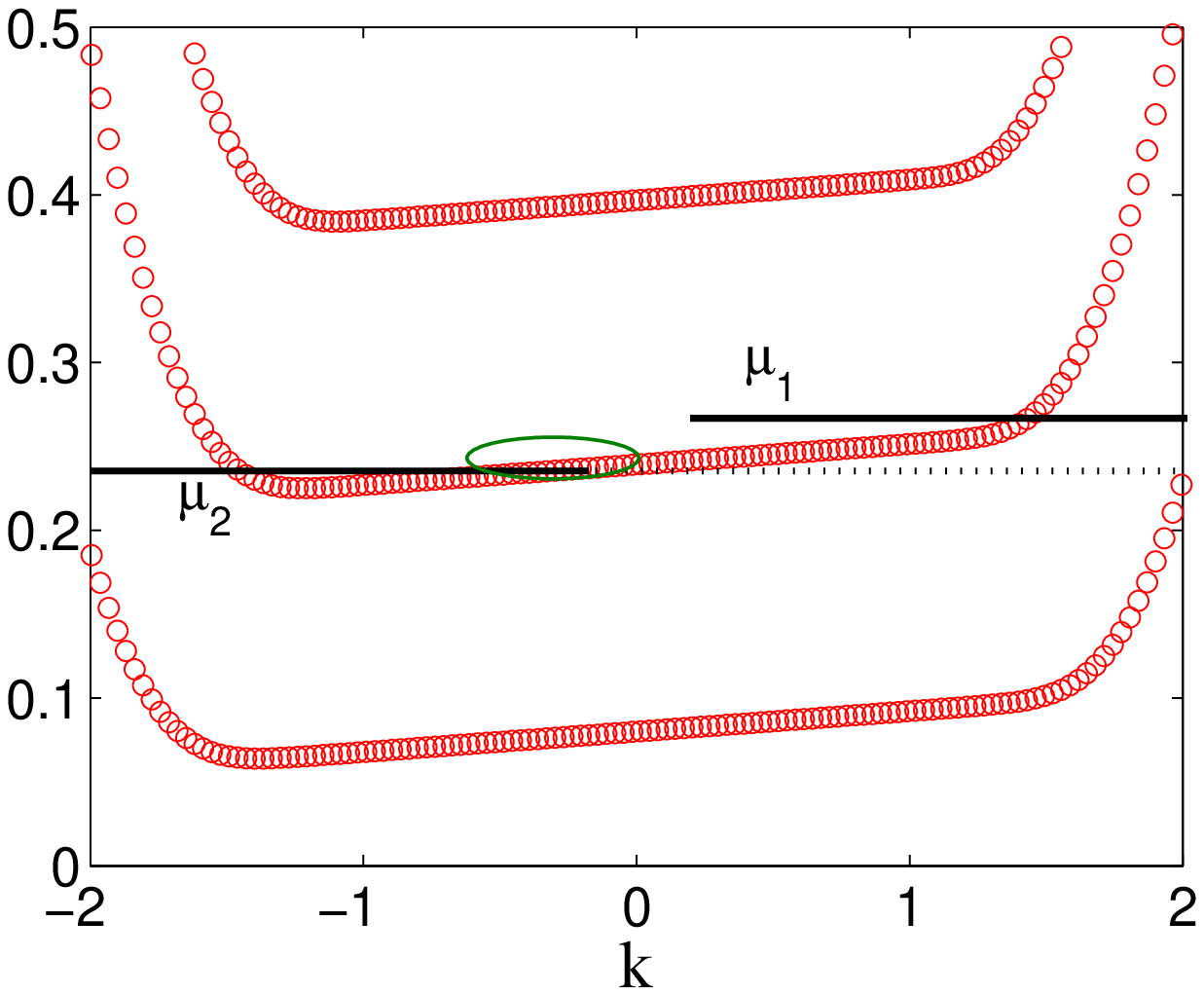}\put(-85,110){(a)}\includegraphics[width=1.8in,height=1.8in]{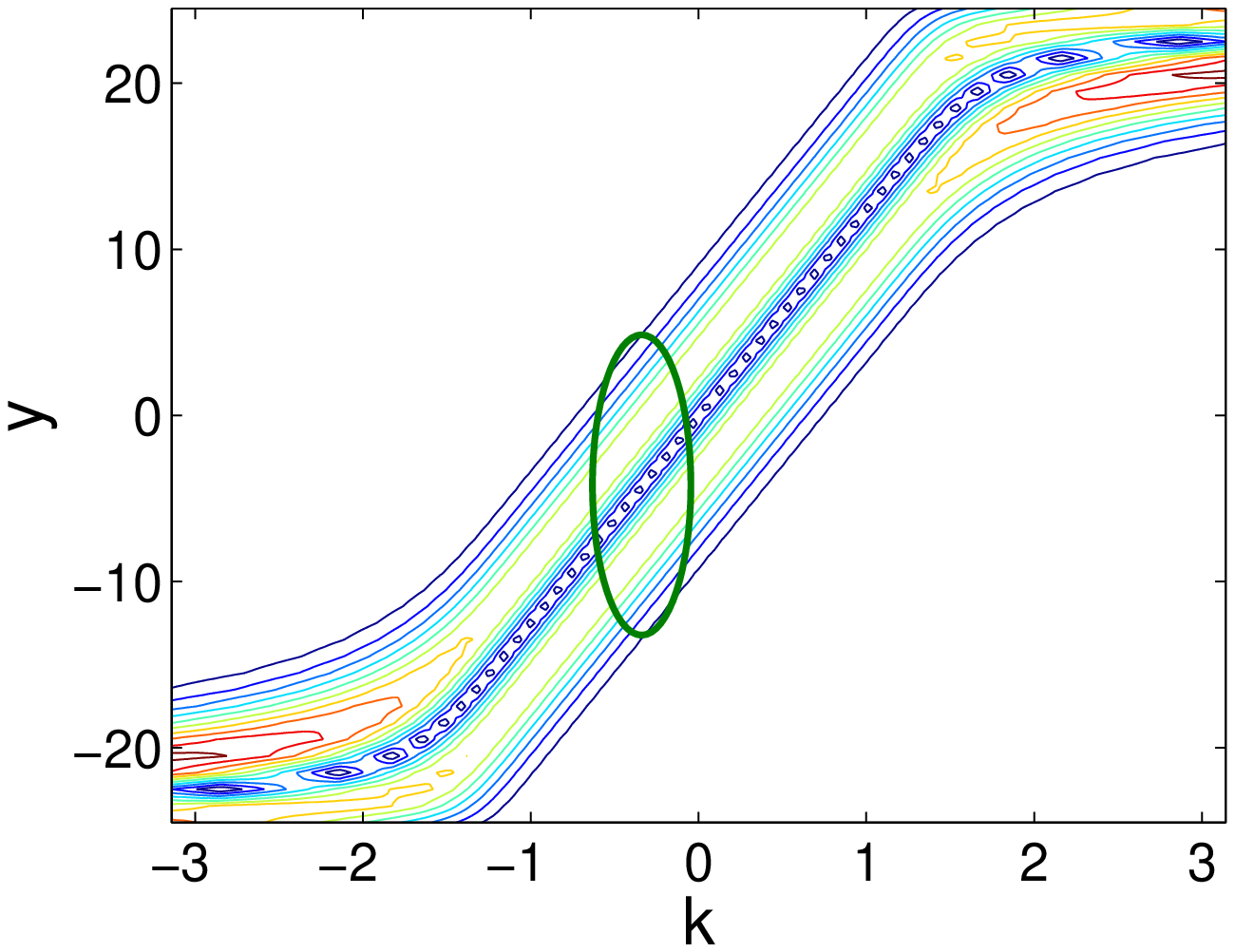}
\put(-100,110){(b)}
\caption{(a)Energies of the lowest Landau levels with the same parameters as in Fig.1 but $\alpha =0.0128$. (b) The modulus of the wave function of the second Landau level. The green circles in both diagrams denote the empty states above the Fermi level in region $k<0$.}
\end{figure}

As shown by the green circles in Fig.3 when the bottom of a Landau level is close to the Fermi level at a certain magnetic field there appears empty states above the Fermi level ($\mu_2$) in the region $k<0$. The states in the region $k>0$ under the Fermi level($\mu_1$),however, are occupied and have higher energies. In addition wave functions of the two regions overlap in space. Thus the electrons in the region $k>0$ easily transfer to the empty states in $k<0$ until a thermal equilibrium. Only when an external voltage between the two ends of the slab is applied a net current may continue. This process causes a heat dissipation and thus a longitudinal resistance. In this case a Landau level provides a bridge for the thermal equilibrium of electrons thus a longitudinal resistance.

According to the theory of Boltzmann equation the conductivity is determined by the relaxation time $\tau_F$ of collisions of electrons on the Fermi surface in the form of $\sigma= ne^2\tau_F/m$. And the relaxation time is given by\cite{Yan}
\begin{align}
 {1\over \tau_F} =\sum_{nn'} \int W_{nkn'k'} (1-\cos\theta_{kk'}) {dk'\over 2\pi}
\end{align}
where $\theta_{kk'}$ is the angle between ${\bf k}$ and ${\bf k}'$ and $W_{nkn'k'}$ is the transition rate. In the present case the electron transport is one dimensional, i.e. $\theta_{kk'}=0,\pi$, but only the back scattering $\theta_{kk'}=\pi$ has contribution to the above integral thus to the longitudinal resistance. The deviation of ions from their lattice sites provides a periodic perturbation to electrons thus causes the transitions of electrons between different electron states.
The Fermi golden rule of quantum mechanics gives
\begin{align}
  W_{nkn'k'} = {2\pi\over \hbar}\sum_{a=\pm 1}|\langle nk|t_a|n'k'\rangle|^2 \delta(E_{nk}-E_{n'k'} -a\hbar \omega)
\end{align}
where $\hbar \omega$ is the energy of phonons and $t_a$ is the perturbation potential between an electron and the lattice vibration.

When the Fermi surface is located between the Landau levels one has $\langle nk|t_a|n'k'\rangle\sim 0 $ since the wave functions of opposite wave vectors are localized at the two edges of the slab as seen in Fig.3(b). This gives an infinite conductivity and thus the longitudinal resistance vanishes at the Hall plateaus. While the Fermi level is close to the bottom of a Landau level the wave functions with opposite and small wave vectors overlap significantly as seen in Fig.3(b) leading to $\langle nk|t_a|n'k'\rangle > 0 $ thus a finite longitudinal resistance appears. Therefore, both the Hall resistance and the longitudinal resistance can be understood in the picture of the electron transport.

\section{conclusions}

The IQHE is analysed using a transport mechanism  with a semi-classic wave packages of electrons in the above sections. A strong magnetic field perpendicular to a slab separates the electron current into two branches with opposite directions, which locate at the two edges of the slab, respectively. This separation prohibits back scattering of electrons ($k\rightarrow -k$). Thus the slab exhibits zero longitudinal resistance and plateaus of Hall resistance. When the Fermi level is scanning over a Landau level when the magnetic field increases, however, the electron waves locate around the central axis of the slab and overlap each other thus back scattering of electrons takes place frequently. Then longitudinal resistance appears and the Hall resistance goes up from one plateau to a new plateau. This mechanism provides a complete explanation to the plateaus of the Hall resistance and the gurgitation of the longitudinal resistance of the IQHE.


\begin{references}
\bibitem{Klitzing} K. von Klitzing {\it et al}, Phys. Rev. Lett.{\bf 45},494(1980).
\bibitem{Tsui} D.C. Tsui {\it et al}, Phys. Rev. Lett.{\bf 48},1559(1982).
\bibitem{Xue} Cui-Zu Chang,Jinsong Zhang, Xiao Feng {\it et al}, Science {\bf 340},167-170(2013).
\bibitem{Prange} R. E. Prange, Phys.Rev.B{\bf 23} (1981)4802.
\bibitem{Ni} Ni Guangjiong and Chen Suqing, Advanced Quantum Mechanics(in chinese),Published by Fudan Univ. Press, p310 (2000).
\bibitem{TKNN} D. J. Thouless, M. Kohmoto, P. Nightingale, and M. den
Nijs, Phys. Rev. Lett. {\bf 49}, 405 (1982); Qian Niu, D. J. Thouless,
and Yong-Shi Wu, Phys. Rev. B{\bf 31}, 3372(1985).
\bibitem{Yasuhiro Hatsugai}Yasuhiro Hatsugai, Phys. Rev. B{\bf 48},11851(1993).
\bibitem{Datta} S. Datta, Electronic Transport in Mesoscopic systems, Cambridge University Press, p181 (1995)
\bibitem{Wees} B.J.van Wees {\it et al.}, Phys. Rev. Lett.{\bf 60},848(1988).
\bibitem{Yan} Yan Shou Sheng, {\it Basis on solid state physics} (in Chinese), Peking University, p126(2003)
\end{references}
\end{document}